# Serf and Turf: Crowdturfing for Fun and Profit

Gang Wang, Christo Wilson, Xiaohan Zhao, Yibo Zhu, Manish Mohanlal,
Haitao Zheng and Ben Y. Zhao
Computer Science, UC Santa Barbara, Santa Barbara, CA 93106, USA
{gangw, bowlin, xiaohanzhao, yibo, manish, htzheng, ravenben}@cs.ucsb.edu

## ABSTRACT

Popular Internet services in recent years have shown that remarkable things can be achieved by harnessing the power of the masses using crowd-sourcing systems. However, crowd-sourcing systems can also pose a real challenge to existing security mechanisms deployed to protect Internet services. Many of these security techniques rely on the assumption that malicious activity is generated automatically by automated programs. Thus they would perform poorly or be easily bypassed when attacks are generated by real users working in a crowd-sourcing system. Through measurements, we have found surprising evidence showing that not only do malicious crowd-sourcing systems exist, but they are rapidly growing in both user base and total revenue. We describe in this paper a significant effort to study and understand these *crowdturfing* systems in today's Internet. We use detailed crawls to extract data about the size and operational structure of these crowdturfing systems. We analyze details of campaigns offered and performed in these sites, and evaluate their end-to-end effectiveness by running active, benign campaigns of our own. Finally, we study and compare the source of workers on crowdturfing sites in different countries. Our results suggest that campaigns on these systems are highly effective at reaching users, and their continuing growth poses a concrete threat to online communities both in the US and elsewhere.

## Categories and Subject Descriptors

H.3.5 [**Information Storage and Retrieval**]: Online Information Services-Web-based services; J.4 [**Computer Applications**]: Social and Behavioral Sciences

## General Terms

Measurement, Security, Economics

## Keywords

Crowdturfing, Crowdsourcing, Spam, Sybils, Experimentation

## 1. INTRODUCTION

Popular Internet services in recent years have shown that remarkable things can be achieved by harnessing the power of the masses. By distributing tasks or questions to large numbers of Internet users, these "crowd-sourcing" systems have done everything from answering user questions (Quora), to translating books, creating 3-D photo tours [29], and predicting the behavior of stock markets and movie grosses. Online services like Amazon's Mechanical Turk, Rent-a-Coder (vWorker), Freelancer, and Innocentive have created open platforms to connect people with jobs and workers willing to perform them for various levels of compensation.

On the other hand, crowd-sourcing systems could pose a serious challenge to a number of security mechanisms deployed to protect Internet services against automated scripts. For example, electronic marketplaces want to prevent scripts from automating auction bids [23], and online social networks (OSNs) want to detect and remove fake users (Sybils) that spread spam [32, 34]. Detection techniques include different types of CAPTCHAs, as well as machine-learning that tries to detect abnormal user behavior [10], *e.g.* near-instantaneous responses to messages or highly bursty user events. Regardless of the specific technique used, they rely on a common assumption, that the malicious tasks in question cannot be performed by real humans en masse. This is an assumption that is easily broken by crowd-sourcing systems dedicated to organizing works to perform malicious tasks.

Through measurements, we have found surprising evidence showing that not only do malicious crowd-sourcing systems exist, but they are rapidly growing in both user base and revenue generated. Because of their similarity with both traditional crowd-sourcing systems and astroturfing behavior, we refer to them as *crowdturfing* systems. More specifically, we define crowdturfing systems as systems where customers initiate "campaigns," and a significant number of users obtain financial compensation in exchange for performing simple "tasks" that go against accepted user policies.

In this paper, we describe a significant effort to study and understand crowdturfing systems in today's Internet. We found significant evidence of these systems in a number of countries, including the US and India, but focus our study on two of the largest crowdturfing systems with readily available data, both of which are hosted in and targeted users in China. From anecdotal evidence, we learn that these systems are well-known to young Internet users in China, and have persisted despite threats from law enforcement agencies to shut them down [5, 9, 20].

Our study results in four key findings on the operation and effectiveness of crowdturfing systems. First, we used detailed crawls to extract data about the size and operational structure of these crowdturfing systems. We use readily available data to quantify both tasks and revenue flowing through these systems, and observe that these sites are growing exponentially in both metrics. Second, we study the types of tasks offered and performed in these sites, which include mass account creation, and posting of specific content on OSNs, microblogs, blogs, and online forums. Tasks often ask users to post advertisements and positive comments about websites along with an URL. We perform detailed analysis of tasks trying to start information cascades on microblogging sites, and study the effectiveness of cascades as a function of the microblog social graph.

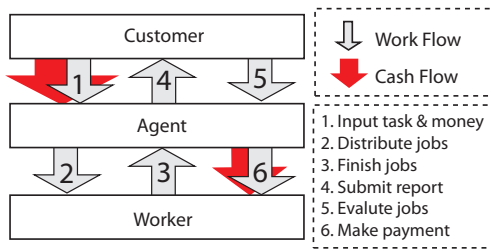

Figure 1: Work and cash flow of a crowdturfing campaign.

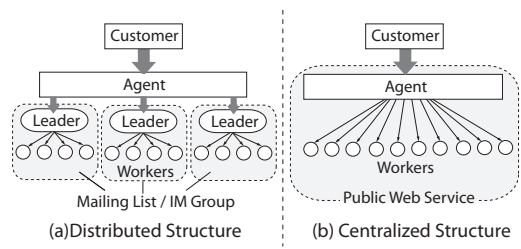

Figure 2: Two different crowdturfing system structures.

Third, we want to evaluate the end-to-end effectiveness of crowdturfing campaigns. To do so, we created accounts on one of our target systems, and initiated a number of benign campaigns that provide unsolicited advertisements for legitimate businesses. By bouncing clicks through our redirection server, we log responses to advertisements generated by our campaigns, allowing us to quantify their effectiveness. Our data shows that crowdturfing campaigns can be cost-effective at soliciting real user responses. Finally, we study and compare the source of workers on crowdturfing sites in different countries. We find that crowdturfing workers easily cross national borders, and workers in less-developed countries often get paid through global payment services for performing tasks affecting US-based networks. This suggests that the continuing growth of crowdturfing systems poses a real threat to U.S.-based online communities such as Facebook, Twitter, and Google+.

This study is the one of the first to examine the organization and effectiveness of large-scale crowdturfing systems on the Internet. These systems have already established roots in other countries, and are responsible for producing fake social network accounts that look indistinguishable from those of real users [34]. A recent study shows that similar types of behavior are also on the rise in the US-based Freelancer site [24]. Understanding the operation of these systems from both financial and technical angles is the first step to developing effective defenses to protect today's online social networks and online communities.

## 2. CROWDTURFING OVERVIEW

In this section, we introduce the core concepts related to crowdturfing. We start by defining crowdturfing and the key players in a crowdturfing campaign. Next, we present two different types of systems that are used to effect crowdturfing campaigns on the Internet: *distributed* and *centralized*. Measurements of a distributed crowdturfing system show that it is significantly less popular with users than centralized systems. Thus we focus on understanding centralized crowdturfing systems in the remainder of our paper.

### 2.1 Introduction to Crowdturfing

The term *crowdturfing* is a portmanteau of "crowd-sourcing" and "astroturfing." Astroturfing refers to information dissemination campaigns that are sponsored by an organization, but are obfuscated so as to appear like spontaneous, decentralized "grass-roots" movements. Astroturfing campaigns often involve spreading legally grey, or even illegal, content, such as defamatory rumors, false advertising, or suspect political messages. Although astroturfing predates the Internet, the ability to quickly mobilize large groups via crowd-sourcing systems has drastically increased the power of astroturfing. We refer to this combined threat as *crowdturfing*. Because of its use of real human users, crowdturfing poses an immediate threat to existing security measures that protect online communities by targeting automated scripts and bots.

Crowdturfing campaigns on the Internet involve three key actors:

1. *Customers:* Individuals or companies who initiate a crowdturfing campaign. The customer is responsible for paying for the monetary costs, and are typically are either related to or themselves the beneficiaries of the campaign.
2. *Agents:* Intermediaries who take charge of campaign planning and management. The agent is responsible for finding, managing, and distributing funds to workers to accomplish the goals of the campaign.
3. *Workers:* Internet users who answer calls by agents to perform specific tasks in exchange for a fee.

Each campaign is structured as a collection of *tasks*. For example, a campaign might entail generating positive sentiment for a new restaurant. In this case, each task would be "post a single (fake) positive restaurant review online." Workers who complete tasks generate *submissions* that include evidence of their work. The customer/agent can then verify that the work was done to their satisfaction. In the case of the restaurant review campaign, submissions are screenshots of or URLs pointing to the fake reviews. Ideally, there is a one-to-one mapping between tasks and submissions. However, not all tasks may be completed, and submissions may be rejected due to lack of quality. In these cases, the number of submissions will not match the number of tasks for a given campaign.

The process for a crowdturfing campaign is shown in Figure 1. Initially, a customer brings the campaign to an agent and pays them to carry it out (1). The agent distributes individual tasks among a pool of workers (2), who complete the tasks and return submissions back to the agent (3). The agent passes the submissions back to the customer (4), who evaluates the work. If the customer is satisfied they inform the agent (5), who then pays the workers (6).

### 2.2 Crowdturfing Systems

*Crowdturfing systems* are instances of infrastructure used to connect customers, agents, and workers to enable crowdturfing campaigns. These systems are generally created and maintained by agents, and help to streamline the process of organizing workers, verifying their work, and distributing payments.

We have observed two different types of crowdturfing systems in the wild: distributed and centralized. We now describe the differences between these two structures, highlighting their respective strengths and weaknesses. Crowdturfing systems are similar to crowd-sourcing systems like Amazon's Mechanical Turk, with the exception that they accept tasks that are unethical or illegal, and that they can utilize distributed infrastructures.

**Distributed Architecture.** Distributed crowdturfing systems are organized around small instant message (IM) groups, mailing lists, or chat rooms hosted by group *leaders*. As illustrated in Fig-ure 2a, leaders act as middlemen between agents and workers, and organizes the workers.

The advantage of distributed crowdturfing systems is that they are resistant to external threats, like law-enforcement. Individual forums and mailing lists are difficult to locate, and they can be dissolved and reconstituted elsewhere at any time. Furthermore, sensitive communications, such as payment transfers, occur via private channels directly between leaders and workers, and thus cannot be observed by third parties.

However, there are two disadvantages to distributed systems that limit their popularity. The first is lack of accountability. Distributed systems do not have robust reputation metrics, leaving customers with little assurance that work will be performed satisfactorily, and workers with no guarantees of getting paid. The second disadvantage stems from the fragmented nature of distributed systems. Prospective workers must locate groups before they can accept jobs, which acts as a barrier-of-entry for many users. To test this, we located 14 crowdturfing groups in China hosted on the popular Tencent QQ instant messaging network. Despite the fact that these groups were well advertised on popular forums, they only hosted ≈2K total users. Over the course of several days of observation, each group only generated 28 messages per day on average, most of which was idle chatter. The conclusion we can draw from these measurements is that distributed crowdturfing systems are not very successful at attracting workers. As we will show in Section 3, centralized systems attract orders of magnitude more campaigns and workers.

**Centralized Architecture.** Centralized crowdturfing systems, illustrated in Figure 2b, are instantiated as websites that directly connect customers and workers. Much like Amazon's Mechanical Turk, customers post campaigns and offer rewards, while workers sign up to complete tasks and collect payments. Both customer and workers register bank information associated with their accounts, and all transactions are processed through the website. Centralized crowdturfing websites use reputation and punishment systems to incentivize customer and workers to behave properly. The primary role of the agent in centralized architectures is simply to maintain the website, although they may also perform verification of submissions at the behest of customers.

The advantage of centralized crowdturfing systems is their simplicity. There are a small number of these large, public websites, making them trivial to locate by customers and workers. Centralized software automates campaign management, payment distribution, and maintains per-worker reputation scores. These features streamline centralized crowdturfing systems, and reduce uncertainties for all involved parties.

The disadvantage of centralized crowdturfing systems is their susceptibility to scrutiny by third parties. Since these public sites allow anyone to sign up, they are easy targets for infiltration, which may be problematic for crowdturfing sites that operate in legally grey-areas. On the other hand, this disadvantage made it possible for us to crawl and analyze several large crowdturfing websites.

## 3. CAMPAIGNS, TASKS, AND REVENUE

We begin our analysis of crowdturfing systems, by analyzing the volume of campaigns, tasks, users, and total revenue processed by the largest known systems. We first describe the representative systems in our study along with our data gathering methodology. We then present detailed results addressing these questions.

### 3.1 Data Collection and General Statistics

While a number of crowdsurfing systems operate across the global Internet, the two largest and most representative systems are hosted on Chinese networks. Their popularity is explained by the fact that China has both the world's largest Internet population (485M) [25] and a moderately low per-capita income (≈$3,200/year) [21]. Crowdturfing sites in China connect dodgy PR firms to a large online user population willing to act as crowd-sourced labor, and have been used to spread false rumors and advertising [4, 20, 5]. This "Shui Jun" (water army), as it is commonly known, has emerged as a force on the Chinese Internet that authorities are only beginning to grapple with [9, 2].

This confluence of factors makes China an ideal place to study crowdturfing. In this section, we measure and characterize the two largest crowdturfing websites in China: Zhubajie (ZBJ, zhubajie.com) and Sandaha (SDH, sandaha.com). All data on these sites are public, and we were able to gather all data on their current and past tasks via periodic crawls of their campaign histories.

**Zhubajie and Sandaha.** The first site we crawled is Zhubajie (ZBJ), which is the largest crowd-sourcing website in China. As shown in Table 1, ZBJ has been active for five years, and is well established in the Chinese market. Customers post many different legitimate types of jobs to ZBJ, including requests for freelance design and programming, as well as Mechanical Turk-style "human intelligence tasks." However, there is a subsection of ZBJ called "Internet Marketing" that is dedicated solely to crowdturfing. ZBJ also has an English-language version hosted in Texas (witmart.com), but its crowdturfing subsection only has 3 campaigns to date. Unlike ZBJ, Sandaha (SDH) only provides crowdturfing services, and is four years younger than ZBJ.

**Crawling Methodology.** We crawled ZBJ and SDH in September, 2011 to gather data for this study. We crawled SDH in its entirety, but only crawled the crowdturfing section of ZBJ. Both sites are structured similarly, starting with a main page that links to a paginated list of campaigns, ordered reverse chronologically. Each campaign has its own page that gives pertinent information, along with links to another paginated list of completed submissions from workers. All information on both sites is publicly available, and neither site employs security measures to prevent crawling.

Our crawler recorded details of all campaigns and submissions on ZBJ and SDH. Campaigns are characterized by a description, start and end times, total number of tasks, total money available to pay workers, whether the campaign is completed, and the number of accepted and rejected submissions. It also includes details for each submission entered by workers, including the worker username and UID, a submission timestamp, one or more screenshots and/or URLs pointing to content generated by the worker, and a flag marking the submission as either accepted or rejected after review.

Both ZBJ and SDH make the complete history of campaigns available on their sites, which enables the crawler to collect data dating back to each site's inception. Table 1 lists the total number of campaigns on each site, as well as the percentage that were usable for our study. Data on some campaigns is incomplete because the customer deleted them or made them private. Other data could be missing because either the campaign only provided partial information (*e.g.* no task count or price per task), or the campaign was still ongoing at the time of our crawl. Incomplete campaigns only account for 8% of the total on ZBJ and 12% on SDH, and thus have little impact on our overall results. For clarity, we convert all currency values on ZBJ and SDH (Chinese Yuan) to US Dollars using an exchange rate of 0.1543 to 1.

**General Statistics.** Table 1 shows the high-level results from our crawls. ZBJ is older and more well-established than SDH, hence it has attracted more campaigns, workers, and money. Campaigns on both sites each include many individual tasks, and task count is almost three orders of magnitude greater than number of campaigns. The number of submissions generated by workers in response to tasks is highly variable: on ZBJ only 36% of tasks receive submissions, whereas on SDH 130% of tasks receive submissions (*i.e.* there is competition among workers to complete the same tasks). Roughly 50% of all submissions are accepted.

| Website | Active Since | Total Campaigns (%) | Total Workers | Total Tasks | Total Submissions (%) | Total Accepted (%) | Total Money | Money for Workers | Money for Website (%) |
|---|---|---|---|---|---|---|---|---|---|
| Zhubajie (ZBJ) | Nov. 2006 | 76K (92%) | 169K | 17.4M | 6.3M (36%) | 3.5M (56%) | $3.0M | $2.4M | $595K (20%) |
| Sandaha (SDH) | March 2010 | 3K (88%) | 11K | 1.1M | 1.4M (130%) | 751K (55%) | $161K | $129K | $32K (20%) |

Table 1: General information for two large crowdturfing websites.

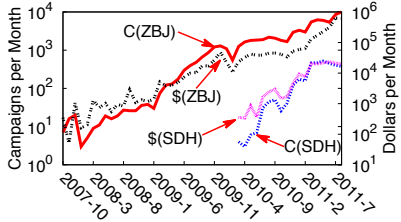
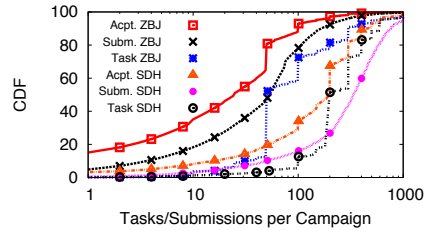
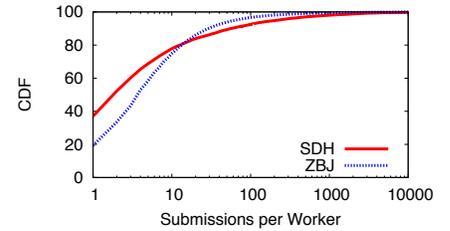

Figure 3: Campaigns, dollars per month.   Figure 4: Tasks, submissions per camp.   Figure 5: Submissions per worker.

Most importantly, more than $4 million dollars have been spent on crowdturfing on ZBJ and SDH in the past five years. Both sites take a 20% cut of campaign dollars as a fee, resulting in significant profits for ZBJ, due to its high volume of campaigns. Furthermore, Figure 3 shows that the number of campaigns and total money spent are growing exponentially. The younger SDH has a growth trend that mirrors ZBJ, suggesting that it will reach similar levels of profitability within the next year. These trends indicate the rising popularity of crowdsurfing systems, and foreshadow the potential impact these systems will have in the very near future.

### 3.2 Campaigns, Tasks, and Workers

Figure 4 illustrates the high level breakdown of tasks and submissions on ZBJ and SDH. There are three lines corresponding to each site: tasks per campaign, submissions per campaign, and accepted submissions per campaign. Campaigns on ZBJ tend to have an order of magnitude fewer tasks than those on SDH. Although both sites only accept ≈50% of submissions, the overabundance of submissions on SDH means that the number of accepted submissions closely tracks the required number of tasks, especially for campaigns with >100 tasks.

**Campaign Types.** Crowdturfing campaigns on ZBJ and SDH can be divided into several categories, with the five most popular listed in Table 2. These five campaign types account for 88% of all campaigns on ZBJ, and 91% on SDH.

"Account registration" refers to the creation of user accounts on a target website. Unlike what has been observed by prior work [24], these accounts are almost never used to automate the process of spamming. Instead, customers request this service to bolster the popularity of fledgling websites and online games, in order to make them appear well trafficked.

Four campaign types refer to spamming in specific contexts: QQ instant-message groups, forums, blogs, and microblogs (*e.g.* Twitter). Customers in China prefer to pay workers directly to generate content on popular websites, rather than purchasing accounts from workers and spamming through them. Note, that QQ and forums represent a larger percentage of campaigns because their existence predates microblogs, which have only become popular in China in the last year [25]. The last column of Table 2 shows the average monthly growth in number of campaigns, and shows that microblogs campaigns are growing faster than all other top-5 categories in both ZBJ and SDH. As the popularity of social networks and microblogs continues to grow, we expect to see more campaigns targeting them.

|  | Campaign Type | Num of Campaigns | $/Camp. | $/Task | Monthly Growth |
|---|---|---|---|---|---|
| ZBJ | Account Reg. | 29,413 (39%) | $71 | $0.35 | 16% |
|  | Forum Post | 17,753 (23%) | $16 | $0.27 | 19% |
|  | QQ Group | 12,969 (17%) | $15 | $0.70 | 17% |
|  | Microblog | 4,061 (5%) | $12 | $0.18 | 47% |
|  | Blog Post | 3,067 (4%) | $12 | $0.23 | 20% |
| SDH | Forum Post | 1,928 (57%) | $48 | $0.19 | 40% |
|  | QQ Group | 473 (14%) | $48 | $0.13 | 31% |
|  | Q&A | 463 (14%) | $47 | $0.21 | 30% |
|  | Blog Post | 113 (3%) | $49 | $0.19 | 21% |
|  | Microblog | 93 (3%) | $49 | $0.27 | 42% |

Table 2: The top five campaign types on ZBJ and SDH.

Finally, "Q&A" involves posting and answering questions on social Q&A sites like Quora (`quora.com`). Workers are expected to answer product-related questions in a biased manner, and in some cases post dummy questions that are immediately answered by other colluding workers.

**Worker Characteristics.** We now focus our discussion on the behavior of workers on crowdturfing websites. Figure 5 shows that the total number of submissions per worker (including both accepted and rejected submissions) varies across the worker population, and even between ZBJ and SDH. Roughly 40% of SDH workers only complete a single task, compared to 20% on ZBJ. The average worker on both sites complete around 5-7 tasks each.

Figure 5 also reveals that a small percentage of extremely prolific workers (especially on SDH) generate hundreds, even thousands, of submissions. Figure 6 plots the percentage of submissions from top workers ordered from most to least prolific. The distribution is highly skewed in favor of these career crowdturfers, who are responsible for generating ≈75% of submissions.

We now examine the temporal aspects of worker behavior. Figure 7 plots the time difference between a campaign getting posted online, and the first submission from a worker. On SDH, 50% of campaigns become active within 24 hours, whereas on ZBJ (with its larger worker population) 75% of campaigns become active within 24 hours. However, some campaigns take significantly longer to ramp up: up to 15 days on ZBJ, and 30 days on SDH. As we discuss in Section 3.3, these slow moving campaigns have very specific requirements that cannot be met by the vast majority of workers.

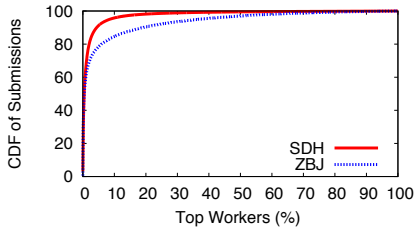
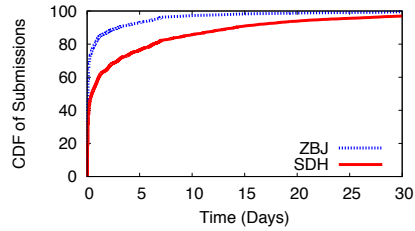
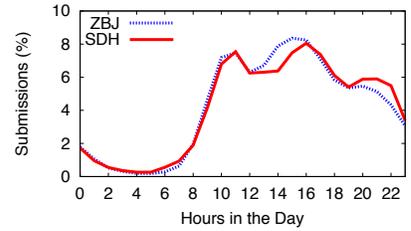

Figure 6: Submissions by top workers.  Figure 7: Time to first response.  Figure 8: Daily submissions.

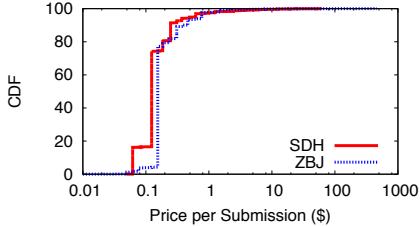
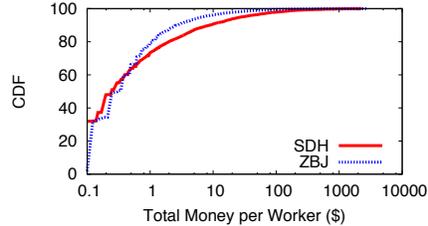
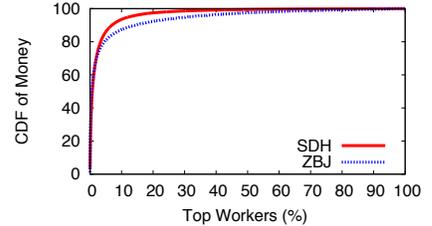

Figure 9: Submission prices.  Figure 10: Money per worker.  Figure 11: Money earned by top workers.

Figure 8 shows the correlation between time of day and number of submissions on ZBJ and SDH. Most submissions happen during the workday and in the evening. Slight drops around lunch and dinner are also visible. This pattern confirms that submissions are generated by human beings, and not automated bots.

### 3.3 Money

We now explore the monetary reward component of crowdturfing systems. As is common on crowd-sourcing systems like Mechanical Turk, workers on ZBJ and SDH make a tiny fee for each accepted submission. As shown in Figure 9, the vast majority of workers on ZBJ and SDH earn $0.11 per submission, although ≈20% of submissions command higher prices than this. Workers must complete many submissions in order to earn substantial pay, leading to the prolific submission habits of career crowdturfers seen in Figure 6. Note that this is a very different model from bid-for-tasks systems like the recent Freelancer study [24].

The total amount of money earned by most workers on ZBJ and SDH is very small. As illustrated in Figure 10, close to 70% of workers earn less than $1 for their efforts. The remaining 30% of workers earn between $1 and $100, making crowdturfing a potentially rewarding part-time job to supplement their core income. For a very small group of workers (0.4%), crowdturfing is a full-time job, earning rewards in the $1,000 dollar range. Not surprisingly, the distribution of monetary rewards matches this distribution. As seen in Figure 11, the top 5% of workers take home 80% of the proceeds on ZBJ and SDH. Clearly, a hard-core contingent of career crowdturfers is taking the bulk of the reward money by quickly completing many submissions.

**Task Pricing.** The goal and budget of each crowdturfing campaign affects the number and price of tasks in that campaign. Figure 12 plots the correlation between the number of tasks in a campaign, versus the price per submission the customer is willing to pay. The vast majority of campaigns with 1K-10K tasks call for generating numerous "tweets" on microblog sites. We examine these tasks in more detail in Section 4.

Although the vast majority of campaigns call for many tasks with low price per submission, Figure 12 reveals that there is a small minority of well paying tasks. In many cases, these campaigns only include a single task that can earn an accepted submission ≥$100 dollars. We examined the 158 outlying tasks that earned ≥$10 and determined that they include a large range of very strange campaigns, some prominent examples include:

- *Pyramid Schemes:* Workers recruit their friends into a pyramid scheme to receive a large payment.
- *Commissioned Sales:* Workers sell products in order to receive a percentage of the sales.
- *Dating Sites:* Workers crawl OSNs and clone the profiles of attractive men and women onto a dating site.
- *Power-Users:* These tasks call for a single worker who owns a powerful social network account, well-read blog, or works for a news service to generate a story endorsing the customer.

## 4. CROWDTURFING ON MICROBLOGS

In this section, we study the broader impact of crowdturfing by measuring the spread of crowdturf content on microblogging sites. We gather data from Sina Weibo, the most popular microblogging social network in China that has the same look and feel as Twitter. We study Weibo for two reasons. First, as shown in Table 2, microblogging sites and Weibo in particular are very popular targets for crowdturfing campaigns. Second, the vast majority of information on Weibo (*i.e.* "tweets" and user profile information) is public, making it an ideal target for measurement and analysis.

We begin by introducing Weibo and our data collection methodology. Next, we examine properties of crowdturfing tasks and workers on Weibo. Finally, we gauge the success of campaigns across the social network by analyzing the spread of crowdturfing content.

### 4.1 Weibo Background and Data Collection

Founded in August 2009, Sina Weibo is the most popular microblogging social network in China, with more than 250 million users as of October 2011 [1]. Weibo has functionality identical to Twitter: users generate 140 character "tweets," which can be replied to and "retweeted" by other users. Users may also create directed relationships with other users by *following* them.

We focus our study of Weibo campaigns from ZBJ, because ZBJ has the most microblogging campaigns by far. Of the 4,061 mi-croblogging campaigns on ZBJ, 3,145 target Weibo. As shown in Figure 13, the number of Weibo campaigns on ZBJ mirrors Weibo's rapid growth in popularity in 2011.

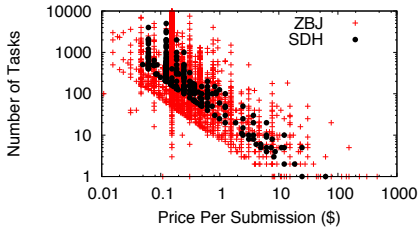
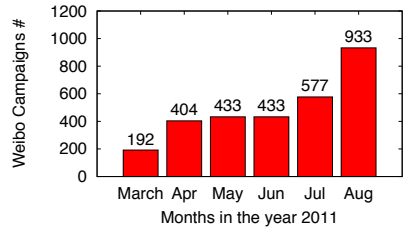
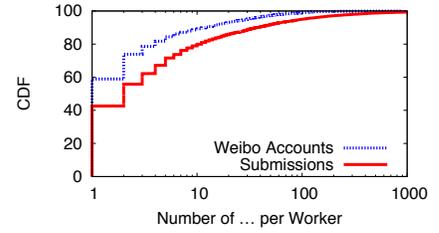

Figure 12: # of tasks vs. submission price.

Figure 13: Weibo campaigns in 2011.

Figure 14: Submissions and accounts.

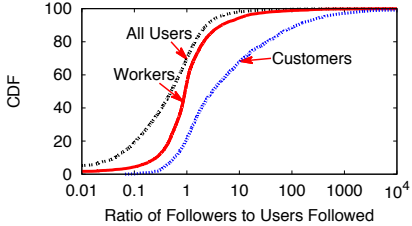
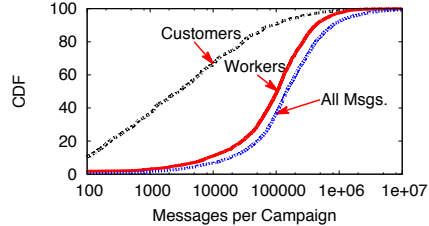
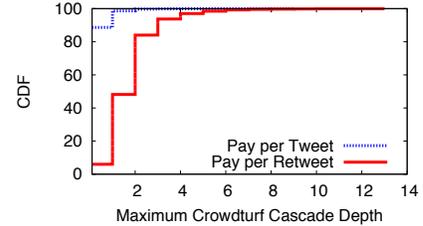

Figure 15: Follow rates for Weibo users.

Figure 16: Messages per campaign.

Figure 17: Crowdturfing cascades.

The goal of crowdturfing campaigns on Weibo is to increase the customer's reach, and to spread their sponsored message throughout the social network. These goals lead to three task types: "pay per tweet," "pay per retweet," and "purchasing followers." The most common task type is retweeting, in which the customer posts a tweet and then pays workers to retweet it. Alternatively, customers may pay workers to generate their own tweets, laden with specific keywords and URLs, or to have their accounts follow the customer's for future messages.

To increase the power of their campaigns, customers prefer workers who use realistic, well-maintained Weibo accounts to complete tasks. Customers may not accept submissions from poor quality, *e.g.* easily detected or banned, Sybil accounts. Conversely, workers who control popular accounts with many followers can earn more per task than worker accounts with average popularity.

**Data Collection.** Understanding the spread of crowdturfing content on Weibo requires identifying *information cascades* [19]. Each cascade is characterized by an *origin post* that initiates the cascade, and retweets that further propagate the information. Cascades form a directed tree with the origin post at the root. In *crowdturfing cascades*, the origin post is always generated by a customer or a worker, but retweets can be attributed to workers and normal Weibo users. Each campaign is a forest of cascade trees.

We crawled Weibo in early September, 2011 to gather data on the spread of crowdturf content. The crawler was initially seeded with URLs that matched campaigns already found on ZBJ, and used simple content analysis to determine if each worker submission was an origin post or a retweet in order to differentiate between "pay to tweet" and "pay to retweet" tasks. In the latter case, the crawler fetched the origin post using information embedded in the retweet.

Our crawler targets the mobile version of the Weibo site because it lists all retweets of a given origin post on a single page, including the full path of multi-hop retweets. The crawler recorded the total number of tweets, followers, and users followed by each user involved in crowdturfing cascades. Unfortunately, Weibo only divulges the first 1K followers for each user, so we are unable to fully reconstruct the social graph.

Overall, our crawler collected 2,869 campaigns involving 1,280 customers. These campaigns received submissions from more than 12,000 Weibo accounts, and reached more than 463,000 non-worker users. Among these, 2% of worker accounts were inaccessible, and were presumably banned by Weibo for spamming. 0.08% of the non-worker user accounts were inaccessible, and all customer accounts remained active. "Pay per tweet" campaigns initiated 25,000 cascades, while "pay per retweet" campaigns triggered 5,000 cascades. We ignore "purchase followers" campaigns, since they do not generate crowdturfing cascades.

To get a baseline understanding of normal Weibo user accounts, we performed a snowball crawl of Weibo's social graph in October 2011. The result is profile data for 6 million "normal" Weibo users.

### 4.2 Weibo Account Analysis

We begin by examining and comparing the characteristics of Weibo accounts controlled by workers and customers to those of normal Weibo users. As shown in Figure 14, the number of accounts controlled by each worker follows the same trend as submissions per worker. This is intuitive: workers need multiple accounts in order to make multiple submissions to a single campaign. Hence, professional crowdturfers who generate many submissions need to control a commensurate number of accounts. In absolute terms, we observe 14,151 accounts controlled by 5,364 ZBJ workers. The top 1% of workers each control $\geq 100$ accounts, but the average worker controls only $\approx 6$ accounts.

**Comparison to Normal Accounts.** We now compare characteristics of worker's and customer's accounts to normal users. We find that each account type tweets with the same frequency. This suggests that workers and customers are both careful not to overwhelm their followers with spam tweets.

Previous work on Sybil detection on OSNs showed that *follow rate* is an effective metric for locating aberrant accounts [31]. A user's follow rate is defined as the ratio of followers to users followed. Sybils often attempt to gain followers by following many other users and hoping they reciprocate. Thus Sybils have follow rates $<1$, *e.g.* they follow more users than they have followers.

Figure 15 shows the follow rates for different Weibo account types. Surprisingly, normal users have the lowest follow rates. Most worker accounts have follow rates $\approx 1$, allowing them to eas-ily blend in. This may represent a conscious effort on the part of workers to make their Weibo accounts appear "normal" so that they will evade automatic Sybil detectors. Customers tend to have follow rates >1. This makes sense, since customers tend to be commercial entities, and are thus net information disseminators rather than information consumers.

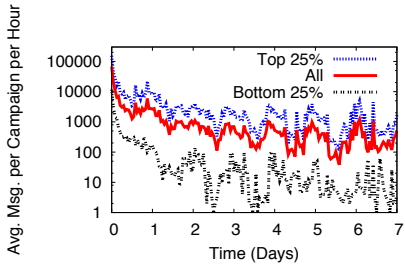
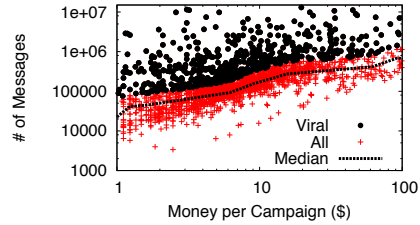
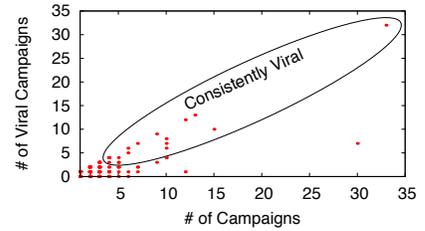

**Figure 18: Message creation over time.**   **Figure 19: Cost of Weibo campaigns.**   **Figure 20: Viral campaigns per customer.**

## 4.3 Information Dissemination on Weibo

Much work has studied how to optimize information dissemination on social networks. We analyze our data to evaluate the level of success in crowdturfing cascades, and whether there are factors that can predict the success of social crowdturfing campaigns.

**Campaign Analysis.** We start by examining the number of *messages* generated by crowdturfing campaigns on Weibo. We define a message as a single entry in a Weibo timeline. A tweet from a single user generates $f$ messages, where $f$ is their number of followers. The number of messages in a campaign is equal to the number of messages generated by the customer, workers, and any normal users who retweet the content. Total messages per campaign represents an upper bound on the *audience size* of that campaign. Since we have an incomplete view of the Weibo social graph, we cannot quantify the number of duplicate messages per user.

Figure 16 shows the CDF of messages generated by Weibo campaigns. 50% of campaigns generate ≤146K messages, and 8% manage to breach the 1M-message milestone. As expected, workers are responsible for the vast majority of messages, *i.e.* there are very few retweets. Considering the low cost of these campaigns, however, these raw numbers are nonetheless impressive.

Next, we want to examine the depth of crowdturfing cascades. Figure 17 plots the depth of cascades measured as the height of each information cascade tree. Pay-per-tweet campaigns are very shallow, *i.e.* worker's tweets are rarely retweeted by normal users. In contrast, pay-per-retweet campaigns are more successful at engaging normal users: 50% reach depths >2, *i.e.* they include at least one retweet from a normal user. One possible explanation for the success of pay per retweet is that normal users may place greater trust in information that is retweeted from a popular customer, rather than content authored by random worker accounts.

Next, we examine the temporal dynamics of crowdturfing campaigns. Figure 18 shows the number of messages generated per hour after each campaign is initiated. The "all" line is averaged across all campaigns, while the top- and bottom-25% lines focus on the largest and smallest campaigns (in terms of total messages). Most messages are generated during a campaigns' first hour (10K on average), which is bolstered by the high-degree of customers (who tend to be super-nodes), and the quick responses of career crowdturfers (see Figure 7). However, by the end of the first day, the message rate drops to ≈1K per hour. There is a two order of magnitude difference between the effectiveness of the top- and bottom-25% campaigns, although they both follow the same falloff trend after day 1.

**Factors Impacting Campaign Success.** We now take a look at factors that may affect the performance of crowdturfing cascades. The high-level question we wish to answer is: are there specific ways to improve the probability that a campaign goes *viral*?

The first factor we examine is the cost of the campaign. Figure 19 illustrates the number of messages generated by Weibo campaigns versus their cost. The median line, around which the bulk of campaigns are clustered, reveals a linear relationship between money and messages. This result is intuitive: more money buys more workers, who in turn generate more messages. However, Figure 19 also reveals the presence of *viral* campaigns, which we define as campaigns that generate at least two times more messages than their cost would predict. There are 723 viral campaigns scattered randomly throughout the upper portion of Figure 19. This shows that viral popularity is independent of campaign budget.

We look at whether specific workers are better at generating viral campaigns. We found that individual workers are not responsible for the success of viral campaigns. The only workers consistently involved in viral campaigns are career crowdturfers, who tend to be involved in *all* campaigns, viral or not.

Surprisingly, a small number of customers exhibit a consistent ability to start viral campaigns. Figure 20 plots the total number of campaigns started by each customer vs. the number that went viral, for all customers who started at least 1 viral campaign. The vast majority of customers initiate ≤3 campaigns, which makes it difficult to claim correlation when one or more go viral. However, the 20 customers (1.5%) in the highlighted region do initiate a significant number of campaigns, and they go viral ≥50% of the time. Since many of these customers do not actively participate in their own campaigns, this suggests that campaigns go viral because their content is of interest to Weibo users, perhaps because they are related to customers such as well-known actors or performers.

## 5. ACTIVE EXPERIMENTS

Our next step to understanding crowdsurfing systems involves a look from the perspective of a paying customer on ZBJ. We initiate a number of benign advertising campaigns on different platforms and subjects. By redirecting the click traffic through a *measurement server* under our control, we are able to analyze the clicks of workers and of users receiving crowdturf content in real-time. We begin by describing our experimental setup before moving on to our findings, and conclude with a discussion of practical lessons we learned during this process.

### 5.1 Experimental Setup

**Methodology.** Figure 21 depicts the procedure we use to collect real-time data on crowdturfing clicks. The process begins when we post a new campaign to ZBJ that contains a brief description of the tasks, along with a URL ("Task Info" in Figure 21) that workers can click on to find details and to perform the tasks. The task details page is hosted on our measurement server, and thus any worker who wants to accept our tasks must first visit our server, where we collect their information (*i.e.* IP, timestamp, etc). Referring workers to task details on external sites is a common practice on ZBJ, and does not raise suspicion among workers.

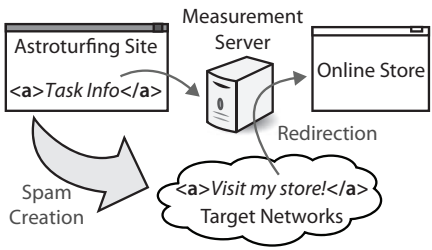
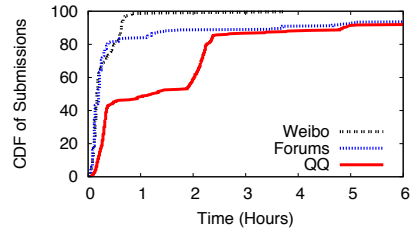
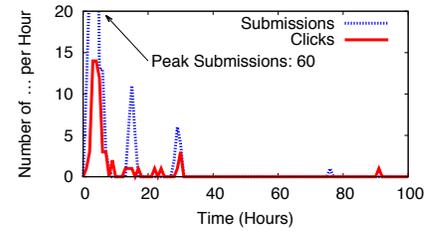

**Figure 21: Crowdturfing data collection.**  **Figure 22: Response time of ZBJ workers.**  **Figure 23: Long campaign characteristics.**

Workers that accept our tasks are directed to post spam messages that advertise real online stores to one of three target networks: Weibo, QQ instant message groups, and discussion forums. The posted messages urge normal users to click embedded links ("Visit my store!" in Figure 21) that take them to our measurement server. The measurement server records some user data before transparently redirecting them to the real online store.

We took care to preserve the integrity of our experimental setup. Because some Chinese Internet users have limited access to websites hosted outside of mainland China, we placed our measurement server in China, and only advertised legitimate Chinese e-commerce sites. In addition, we also identified many search engines and bots generating clicks on our links, and filtered them out before analyzing our logs.

**Campaign Details.** In order to experiment with a variety of topics and venues, we posted nine total campaigns to ZBJ in October 2011. As shown in Table 3, we created three different advertising campaigns (*iPhone4S*, *Maldives*, and *Raffle*), and targeted each at three distinct networks. We discuss a fourth campaign, *OceanPark*, later in the section.

The first campaign promotes an unofficial iPhone dealer who imports iPhones from North America and sells them in China. We launched this campaign on October 4, 2011, immediately after Apple officially unveiled the iPhone 4S. In the task requirements, we required workers to post messages advertising a discount price from the dealer on the iPhone 4S ($970).

The second campaign tried to sell a tour package to the Maldives (a popular tourist destination in China). The spam advertises a 30% group-purchase discount offered by the seller that saves $600 on the total trip price ($1542 after discount). The third campaign tells users about an online raffle hosted by a car company. Anyone could participate in the raffle for free, and the prizes were 200 pre-paid calling cards worth $4.63 each.

All campaigns shared the same set of baseline requirements. Each campaign had a budget of $15 on each target network, and workers had a time limit of 7 days to perform tasks. The desired number of tasks was set to either 50 or 100, depending on the campaign type. Submissions were not accepted if the content generated by the worker was deleted by spam detection systems within 24 hours of creation. These baseline requirements closely match the expected norms for campaigns on ZBJ (see Figure 4 and Table 2).

We applied additional requirements for campaigns on specific networks. For campaigns on the QQ instant messaging network, workers were required to generate content in groups with a minimum of 300 members. For campaigns on user discussion forums, workers were only allowed to post content on a predefined list of forums that receive at least 1,000 hits per day.

Each campaign type had additional, variable requirements. For Maldives and Raffle campaigns, the price per task was set to $0.154, meaning 100 submissions would be accepted. However, the price for iPhone4S tasks was doubled to $0.308 with an expectation of 50 submissions. iPhone 4S tasks were more challenging for two reasons. On Weibo, workers were required to tweet using accounts with at least 3,000 followers. On QQ, workers needed to spam two groups instead of one. Finally, on forums, the list of acceptable sites was reduced to only include the most popular forums.

## 5.2 Results and Analysis

Table 3 lists the high level results of from our crowdturfing campaigns, including 9 short campaigns and the "OceanPark" campaign. Seven of the short campaigns received sufficient submissions, and six were completed within a few hours (Time column). Interestingly, workers continued submitting to campaigns even after they were "full," in the hopes that earlier submissions would be rejected, and they would claim the reward. In total, the short campaigns garnered 894 submissions from 224 distinct workers.

Figure 22 shows the response times of workers for campaigns targeting different networks. We aggregate the data across campaign types rather than networks because workers' ability to complete tasks is based on the number of accounts they control on each network. More than 80% of submissions are generated within an hour for Weibo and forum campaigns, and within six hours for QQ.

The "Msgs" column lists the number of *messages* generated by each campaign. For Weibo campaigns, we calculate messages using the same methodology as in Section 4. For QQ campaigns, messages are calculated as the number of users in all QQ groups that received spam from our workers. We cannot estimate the number of messages for forums because we do not know how many users browse these sites.

We can understand the effectiveness of different crowdturfing strategies by comparing the number of messages generated to the number of clicks (responses by normal users, "Clicks" column in Table 3). We see that QQ campaigns are the most effective, and generate more clicks than Weibo campaigns despite generating only 1/5 as many messages as Weibo. One possible reason is that QQ messages pop-up directly on users' desktops, leading to more views and clicks. Tweets on Weibo, on the other hand, are not as invasive, and may get lost in the flood of tweets in each user's timeline. Forums perform the worst of the three, most likely because admins on popular forums are diligent about deleting spammy posts.

Finally, we try to detect the presence of Sybil accounts (multiple accounts controlled by one user) on crowdturfing sites. Column "W/IP" in Table 3 compares the number of distinct workers ($W$) to the number of distinct IPs ($IP$) that click on the "Task Info" link (see Figure 21) in each campaign. If $W > IP$, then not all ZBJ workers clicked the link to read the instructions. This suggests that multiple ZBJ worker accounts are controlled by a single user, who viewed the instructions once before completing tasks from multiple accounts. Our results show that $W>IP$ for 66% of our campaigns. Thus, not only do crowdturfers utilize multiple accounts on target websites to complete tasks (Figure 14), but they also have multiple accounts on crowdturfing sites themselves.

| Campaign | Network | Subm. | Time | Msgs. | Clicks | W/IP |
|---|---|---|---|---|---|---|
| iPhone4S | Weibo | 47 | 45min | 197K | 204 | 24/54 |
| | QQ | 41 | 6hr | 35K | 244 | 34/36 |
| | Forums | 71 | 3day | N/A | 43 | 40/22 |
| Maldives | Weibo | 108 | 3h | 220K | 28 | 35/30 |
| | QQ | 118 | 4h | 46K | 187 | 24/29 |
| | Forums | 123 | 4h | N/A | 3 | 18/11 |
| Raffle | Weibo | 131 | 2h | 311K | 47 | 67/38 |
| | QQ | 131 | 6day | 60K | 78 | 29/33 |
| | Forums | 124 | 1day | N/A | 0 | 28/9 |
| OceanPark | Weibo | 204 | 4day | 369K | 63 | 204/99 |

Table 3: Results from our crowdturfing campaigns.

**Long Campaigns.** The campaigns we have analyzed thus far all required ≤100 tasks, and many were completed within about an hour by workers (see Figure 22). These short campaigns favor career crowdturfers, who control many accounts on target websites and move rapidly to generate submissions.

To observe the actions of less prolific workers, we experimented with a longer campaign that required 300 tasks. This campaign included an additional restriction to limit career crowdturfers: each ZBJ worker account could only submit once. The goal of the campaign was to advertise discount tickets to an ocean-themed amusement park in Hong Kong on Weibo. This campaign is listed as *OceanPark* in Table 3.

Figure 23 plots the number of worker submissions and clicks from Weibo users over time for the OceanPark campaign. Just as in previous experiments, the first 100 submissions were generated within the first few hours. Clicks from users on the advertised links closely track worker submission patterns. Overall, 191 submissions were received on day one, 11 more on day two, and 2 final submissions on day four, for a total of 204 submissions. This indicates that there are ≈200 active Weibo workers on ZBJ: if there were more, they would have submitted to claim one of the 97 incomplete tasks in our campaign.

**Discussion.** Our real-world experiments demonstrate the feasibility of crowd-sourced spamming. The iPhone4S and Maldives campaigns were able to generate 491 and 218 click-backs (respectively) while only costing $45 each. Considering that the iPhone 4S sells for $970 in China, and the Maldives tour package costs $1,542, just a single sale of either item would be more than enough to recoup the entire crowdturfing fee. The *cost per click* (CPC) of these campaigns are $0.21 and $0.09, respectively, which is more expensive than observed CPC rates ($0.01) for traditional display advertising on the web [30]. However, with improved targeting (*i.e.* omitting underperformers like forum spam) the costs could be reduced, bringing CPC more in line with display advertising.

Our Maldives campaign is a good indicator of the effectiveness of crowdturfing. The tour website listed 4 Maldives trips sold to 2 people in the month before our campaign. However, the day our Maldives campaign went live, 11 trips were sold to 2 people. In the month after our campaign, no additional trips were sold. While we cannot be sure, it is likely that the 218 clicks from our campaign were responsible for these sales.

## 6. CROWDTURFING GOES GLOBAL

In previous sections, we focused on the crowdturfing market in China. We now take a global view and survey the market for crowd-turfing systems in the U.S. and India. Additional crawls conducted by us, as well as prior work from other researchers, demonstrates that crowdturfing systems in the U.S. are very active, and are supported by an international workforce.

| Website | Campaigns | % Crowd-turfing | Tasks | $ per Subm. |
|---|---|---|---|---|
| Amazon Turk (US) | 41K | 12% | 2.9M | $0.092 |
| ShortTask* (US) | 30K | 95% | 527K | $0.096 |
| MinuteWorkers (US) | 710 | 70% | 10K | $0.241 |
| MyEasyTask (US) | 166 | 83% | 4K | $0.149 |
| Microworkers (US) | 267 | 89% | 84K | $0.175 |
| Paisalive (India) | 107 | N/A | N/A | $0.01 |

Table 4: Details of U.S. and Indian crowd-sourcing sites. Data encompasses one month of campaigns, except ShortTask which is one year.

**Mechanical Turk.** Although prior work has found that 41% of tasks on Mechanical Turk were spam related in 2010 [15], our measurements indicate that this is no longer the case. We performed hourly crawls of Mechanical Turk for one month in October 2011, and used keyword analysis to classify tasks. As shown in Table 4, crowdturfing now only accounts for only 12% of campaigns.

**Other U.S. Based Sites.** However, the drop in crowdturfing on Mechanical Turk does not mean this problem has gone away. Instead, crowdturfing has just shifted to alternative websites. For example, recent work has shown that 31% of the jobs on Freelancer over the last seven years were related to search engine optimization (SEO), Sybil account creation, and spam [24]. Many SEO products are also available on eBay: trivial keyword searches turn up many sellers offering bulk Facebook likes/fans and Twitter followers.

To confirm this finding, we crawled four U.S. based crowd-sourcing sites that have been active since 2009. Since they do not provide information on past tasks, we crawled MinuteWorkers, MyEasyTask, and Microworkers once a day during the month of October 2011. ShortTask does provide historical data for tasks going back one year, hence we only crawled them once. As shown in Table 4, keyword classification reveals that between 70-95% of campaigns on these sites are crowdturfing. We manually verified that the remaining campaigns were not malicious. The types of campaigns on these sites closely matches the types found on Freelancer, *i.e.* the most prevalent campaign type is SEO [24].

Sites like ShortTask, Microworkers, and MyEasyTask fill two needs in the underground market. First, they do not enforce any restrictions against crowdturfing. This contrasts with Mechanical Turk, which actively enforces policies against spammy jobs [6]. Second, these sites enable a truly international workforce by supporting a wide range of payment methods. Amazon requires workers to have U.S. bank accounts, or to accept cheques in Indian Rupees, and hence most "turkers" are located in the US (46.8%) and India (34%) [14]. However, alternative crowd-sourcing sites support payments through systems like Paypal and E-Gold, which makes them accessible to non-U.S. and non-Indian workers. For example, Microworkers come from Indonesia (18%), Bangladesh (17%), Philippines (5%), and Romania (5%) [12]. Freelancers are also located in the United Kingdom and Pakistan [24].

**Paisalive.** We located one crowdturfing site in India called Paisalive that takes globalization even further. As shown in Table 4, Paisalive is very small and the wages are very low compared to other services. However, the interesting feature of Paisalive is that it is e-mail based: workers sign up on the website, and afterwards all task requests and submissions are handled through e-mail. This design is geared towards enabling workers in rural populations constrained by low-bandwidth, intermittent Internet connectivity.

## 7. RELATED WORK

**Crowd-sourcing Research.** Since coming online in 2005, Amazon's Mechanical Turk has been scrutinized by the research community. This includes studies of worker demographics [14, 28], task pricing [7, 13], and even meta-studies on how to use Mechanical Turk to conduct user studies [18]. The characteristics of Micro Workers have also been thoroughly studied [12].

**OSN Spam and Detection.** Researchers have identified copious amounts of fake accounts and spam campaigns on large OSNs like Facebook [8], Twitter [11, 32], and Renren [34]. The growing threat posed by this malicious activity has spurred work that aims to detect and stop OSN spam using machine learning techniques [3, 33, 31]. This body of research has focused on analyzing and defending against the outward manifestations of OSN spam. In contrast, our work identifies some of the underlying systems used by attackers to generate spam and evade security measures.

**Opinion Spam.** Spam that attempts to influence the opinions and actions of normal people has become more prevalent in recent years [16]. Researchers have been working on detecting and characterizing fake product reviews [22, 17], fake comments on news sites [4], and astroturf political campaigns on Twitter [27]. The authors of [26] created a model to help classify deceptive reviews generated by Mechanical Turk workers. These works reaffirm our results, that crowdturfing is a growing, global threat on the web.

## 8. CONCLUSION

In this paper, we contribute to the growing pool of knowledge about malicious crowd-sourcing systems. Our analysis of the two largest crowdturfing sites in China reveals that $4 million dollars have already been spent on these two sites alone. The number of campaigns and dollars spent on ZBJ and SDH are growing exponentially, meaning that the problems associated with crowdturfing will continue to get worse in the future.

We measure the real-world ramifications of crowdturfing by looking at spam dissemination on Weibo, and by becoming active customers of ZBJ. Our results reveal the presence of career crowdturfers that control thousands of accounts on OSNs, and manage them carefully by hand. We find that these workers are capable of generating large information cascades, while avoiding the security systems that are designed to catch automated spam. We also observe that this spam is highly effective, driving hundreds of clicks from normal users.

Finally, our survey of crowdturfing sites in the U.S. and elsewhere demonstrates the global nature of this problem. Unscrupulous crowd-sourcing sites, coupled with international payment systems, have enabled a burgeoning crowdturfing market that targets U.S. websites, fueled by a global workforce. As part of ongoing work, we are exploring the design and quantifying the effectiveness of both passive and active defenses against these systems.

## Acknowledgments

The authors wish to thank Zengbin Zhang for his help on ZBJ experiments. This project is supported by NSF under IIS-0916307, and also supported by WCU (World Class University) program under the National Research Foundation of Korea and funded by the Ministry of Eduation, Science and Technology of Korea (Project No: R31-30007).